\documentclass[amsmath,amssymb,twocolumn,amsfonts]{revtex4-2}
\usepackage{graphicx}
\usepackage{epsfig}
\usepackage{bm}
\usepackage{bbm,mathbbol}
\usepackage{braket,bigints}
\usepackage{lmodern}
\usepackage{amsthm,mathtools,bm}
\usepackage{physics}
\usepackage{siunitx}
\usepackage{enumitem}
\usepackage{hyperref}
\usepackage{stmaryrd,mathtools}
\usepackage{csquotes}
\hypersetup{colorlinks=true,linkcolor=blue,citecolor=blue,urlcolor=blue}

\def \beq {\begin{equation}}
\def \eeq {\end{equation}}

\begin{document}
\title{Searching for quantum effects in the brain: A Bell-type test for nonclassical latent representations in autoencoders}
\author{I. K. Kominis}
\email{ikominis@uoc.gr}
\affiliation{$^1$School of Science, Zhejiang University of Science and Technology, Hangzhou 310023, People’s Republic of China}
\affiliation{$^2$Zhejiang Key Laboratory of Biomedical Intelligent Computing Technology, Hangzhou 310023, People’s Republic of China}
\affiliation{$^3$Department of Physics and Institute of Theoretical and Computational Physics, University of Crete, Heraklion 70013, Greece}
\author{C. Xie}
\affiliation{$^4$College of Life and Environmental Sciences, Hangzhou Normal University, Hangzhou 311121, China}
\author{S. Li}
\affiliation{$^5$School of Automation and Electrical Engineering, Zhejiang University of Science and Technology, Hangzhou 310023, China}
\author{M. Skotiniotis}
\affiliation{$^6$Quantum Thermodynamics and Computation Group, Departamento de Electromagnetismo y F\'isica de la Materia, Universidad de Granada, 18071 Granada, Spain}
\affiliation{$^7$Instituto Carlos I de F\'isica Te\'orica y Computacional, Universidad de Granada, 18071 Granada, Spain}
\author{G. P. Tsironis}
\affiliation{$^3$Department of Physics and Institute of Theoretical and Computational Physics, University of Crete, Heraklion 70013, Greece}
\affiliation{$^8$John A. Paulson School of Engineering and Applied Sciences, Harvard University, Cambridge, MA 02138, USA}


\begin{abstract}
Whether neural information processing is entirely classical or involves quantum-mechanical elements remains an open question. Here we propose a model-agnostic, information-theoretic test of nonclassicality that bypasses microscopic assumptions and instead probes the structure of neural representations themselves. Using autoencoders as a transparent model system, we introduce a Bell-type consistency test in latent space, and ask whether decoding statistics obtained under multiple readout contexts can be jointly explained by a single positive latent-variable distribution. By shifting the search for quantum-like signatures in neural systems from microscopic dynamics to experimentally testable constraints on information processing, this work opens a new route for probing the fundamental physics of neural computation. The proposed test identifies violations of classical latent-variable consistency at the level of statistical representations, without assuming a specific underlying physical mechanism.
\end{abstract}
\maketitle
\section{Introduction}
The possibility that quantum effects play a functional role in the brain has long been a subject of debate \cite{Hameroff1996}. Much of this discussion emphasized the (non)-sustainability of long-lived quantum coherence or entanglement in neural systems under physiological conditions \cite{Tegmark2000,Hagan2002}. In recent years, related discussions have resurfaced \cite{Adams2020,Gassab2025}, motivated in part by growing evidence that quantum coherence can survive in biological processes, most notably in photosynthetic light harvesting and magnetoreception \cite{Scholes2017,Kominis2015,Kominis2025}, and can confer functional advantages \cite{Plenio2013,Kominis2020}. These developments, together with the extreme multiscale complexity of neural systems and the persistent challenge of explaining consciousness in physical terms \cite{Tegmark2015,Tononi2016,Georgiev2025,Tuszynski2026}, have renewed interest in whether analogous mechanisms could arise in the brain \cite{Fisher2015,Halpern2019,Rourk2021,Manzano2022,Liu2024,Neven2024,Manzano2024}. Nonetheless, readily testable proposals remain scarce. 

Here we depart from the traditional strategy of seeking quantum coherence in specific biophysical substrates. Instead, we adopt an information-theoretic perspective inspired by recent developments in machine learning \cite{Carleo2019,Mehta2019,TsironisBook} and quantum machine learning \cite{Petruccione2015,Biamonte2017,Briegel2018}, shifting the focus from microscopic physical realizations to the structure, efficiency, and cost of information processing itself. The extreme complexity of the brain, together with the vast volume of information it must process under strict energetic constraints \cite{Lennie2003}, strongly suggests the necessity of efficient mechanisms for compression and representation. In machine learning, autoencoders provide a standard framework for formalizing such compression, mapping high-dimensional inputs into a low-dimensional latent space from which the original inputs are reconstructed. The requirement of faithful reconstruction forces the latent variables to distill the essential structure of the input data \cite{Tishby2015}. If neural systems were to exploit quantum effects at all, it would be plausible for them to arise in encoding and compression tasks, where quantum thermodynamic resources might offer advantages to the management of information, energy and entropy bottlenecks \cite{Parrondo2015,Anders2016}.
\begin{figure*}[t]
\begin{center}
\includegraphics[width=15 cm]{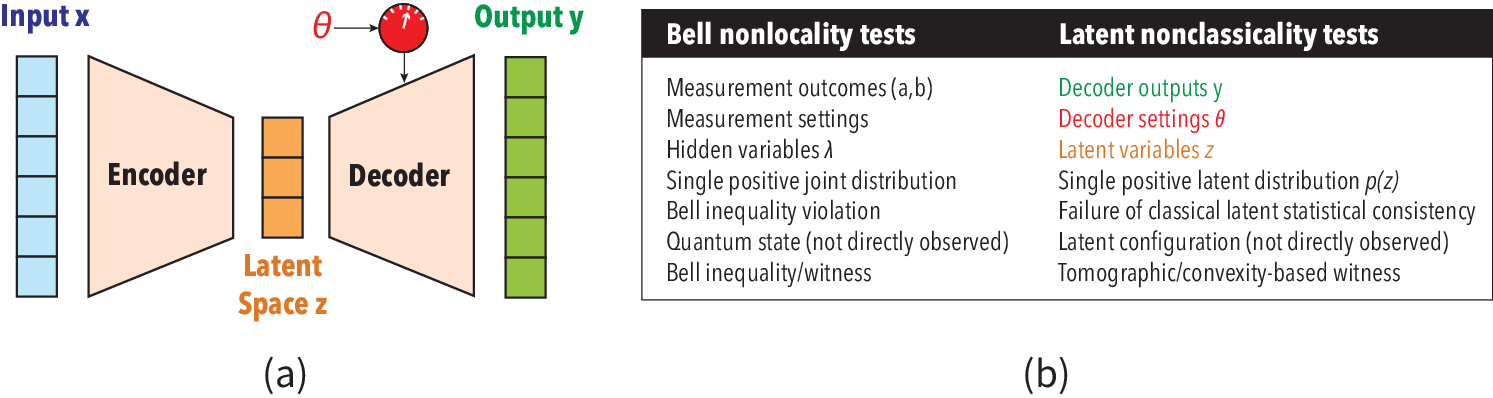}
\caption{(a) An autoencoder maps high-dimensional inputs $x$ to a low-dimensional latent representation and reconstructs outputs $y$ through multiple decoder settings $\theta$. The latent variables are treated as unobserved degrees of freedom probed indirectly via observable decoding marginals. Nonclassicality in the latent space reflects the impossibility to account for all marginal output statistics (different decoder settings $\theta$) with a unique positive latent distribution. (b) Analogy between Bell nonlocality tests and nonclassicality tests in the latent space of autoencoders. }
\end{center}
\end{figure*}

In this work we introduce a purely statistical test of nonclassicality for latent representations learned by autoencoders, and address the question: given a collection of observable reconstruction statistics obtained under different decoder settings, can {\it all corresponding marginals} be realized as projections of a \emph{single positive} latent probability distribution, or is such a joint classical description impossible? Conceptually, we adopt the operational logic of Bell or contextuality tests \cite{Spekkens2008,Cabello2008,Horodecki2009,Brunner2014,Shalm2015,Cavalcanti2017,Watts2021,Budroni2022}, with decoder settings $\theta$ playing the role of measurement contexts and latent variables $z$ acting as hidden variables (Fig.~1). The assumption of a single context-independent latent distribution is the analogue of measurement independence/noncontextuality in Bell-type scenarios: changing the readout context probes different aspects of the same underlying representation rather than preparing a different latent state for each context. 

We note for completeness that theory-agnostic approaches have also been explored in broader quantum-biology contexts, where one asks whether experimentally observed correlations require nonclassical descriptions without committing to a specific microscopic mechanism \cite{Wilde2010,YungerHalpern2020,Spaventa2025}. Moreover, there are numerous works on quantum autoencoders, which posit intrinsically quantum latent subsystems by construction \cite{Guzik2017,Prieto2021,Sakhnenko2022,Locher2023,Tsironis2023,Ma2024,Wu2025,Frehner2025}. Instead, the present work provides a Bell-type hypothesis test for the adequacy of classical latent-variable explanations themselves, entirely independent of any assumed quantum implementation.

In more detail, consider data samples $x$ drawn from a distribution $p_{\mathrm{data}}(x)$, an encoder $E:x\mapsto z$, and a decoder $D:z\mapsto y$. The encoder maps high-dimensional inputs $x$ into a typically lower-dimensional latent variable $z$, which is intended to capture a compressed representation of the data sufficient for reconstruction. A generic autoencoder is trained to minimize a reconstruction error of the form $\mathbb{E}_x[\ell(x,D(E(x)))]$, where $\ell(x,y)$ is a loss function. While any fixed encoder $E$ induces a conditional distribution $p(z\mid x)$, the reconstruction objective generally admits many distinct encoders, and hence many distinct latent representations, achieving comparable loss. As a result, the associated aggregate latent distribution, $p(z)=\int p(z\mid x)\,p_{\mathrm{data}}(x)\,dx$, is not uniquely specified by the learning objective. This underdetermination motivates the question of which additional, ensemble-level restrictions on latent representations can be meaningfully imposed and, crucially, tested against observable data.

A seemingly innocuous such restriction is the assumption that all observable statistics {\it across readout contexts} (to be defined shortly) arise from a {\it single positive} latent distribution. Although this assumption appears natural and is widely taken for granted, it constitutes a foundational constraint that we here scrutinize. In the present setting, this assumption has an operational rather than microscopic meaning. It states that changing the readout context probes different aspects of the same underlying representation, rather than preparing a different latent ensemble for each context. In this sense, the context-independent distribution \(p(z)\) plays a role analogous to the hidden-variable distribution in Bell or contextuality tests, while the decoder setting \(\theta\) plays the role of the measurement context. A violation of this assumption should therefore be interpreted as evidence against a single positive latent-variable description of the observed statistics.
\section{Non-classicality witness in autoencoder statistics}
Concretely, in a classical latent-variable model the conditional distribution of observable outputs $y$, given an externally specified readout context $\theta$, is assumed to arise from a \emph{single positive latent distribution} $p(z)\geq 0$, according to
\begin{equation}
p(y\mid\theta)=\int dz\,p(y\mid z,\theta)\,p(z)
\label{eq:classical_latent}
\end{equation}
As in contextuality tests, classicality is not a property of statistics obtained in a single context, but of their joint consistency across multiple contexts with one underlying latent distribution. 

Operationally, experiments provide access to samples of the outputs $y$ for each context $\theta$, from which the conditional distributions $p(y\mid\theta)$ can be estimated. To formulate a finite-dimensional consistency test, we consider the probabilities \( p(y_k \mid \theta_j) \) of observing outcome \( y_k \) under measurement context \( \theta_j \), with \( j = 1,\dots,J \) and \( k = 1,\dots,K \). The $J\times K$ array containing the probabilities $p(y_k \mid \theta_j)$ is concatenated (flattened) into a single vector $\mathbf p \in \mathbb R^{JK}$. Finally, the space of latent distributions is also discretized, yielding a coefficient vector $\mathbf w \in \mathbb R^{N}$, with $\sum_i w_i = 1$, and where $N$ is independent of the number of measured contexts or outcomes.

The decoder induces a linear map on latent distributions, represented in the finite-dimensional setting as $\mathbf p = A\,\mathbf w$, where the matrix $A \in \mathbb R^{(J K)\times N}$ is fully determined by the decoder and the chosen set of readout settings $\{\theta_j\}$. Operationally, the matrix \(A\) encodes how latent-space regions contribute to observable outputs across different readout contexts. In a data-driven implementation, the entries of \(A\) may be estimated empirically from conditional frequencies associated with latent-space bins and decoder settings, without assuming an underlying microscopic model. More generally, the framework applies whenever one can identify: (i) a latent representation space, (ii) multiple readout or decoding contexts, and (iii) observable conditional output statistics. The same formalism may therefore be implemented not only for phase-space models, but also for neural recordings, learned autoencoder representations, recurrent neural-network activations, or other high-dimensional statistical systems.

We note for completeness how the forward matrix \(A\) is constructed operationally. From the ensemble of observed inputs \(x\), one defines a fixed, low-dimensional representation \(z\), which serves solely as an operational coordinate system for organizing the input data and carries no a priori physical interpretation. The representation space is then partitioned into a finite number \(N\) of regions \(\{z_i\}\).

More explicitly, the entries of the forward matrix are defined as $A_{(j,k),i} \equiv p(y_k \mid z \in z_i, \theta_j)$, i.e. the conditional probability of observing outcome $y_k$ under context $\theta_j$, given that the latent representation lies in region $z_i$. These probabilities are estimated empirically from repeated trials by restricting to samples whose representations fall within the corresponding region. Crucially, for each readout context \(\theta_j\) and each region \(z_i\), these conditional probabilities are estimated from data that are statistically independent from those used to evaluate the observed output statistics $\mathbf p$, ensuring that the test does not suffer from data reuse or overfitting effects. That is, the representation $z$ is introduced only as an operational coordinate system for organizing the data, and is not trained or optimized to reproduce the observed outputs $y$. The proposed test therefore probes the cross-context consistency of the observed decoding statistics, rather than the predictive performance of a learned decoder model.

The set of classically admissible decoding statistics forms the convex polytope $\mathcal C=\{ A\mathbf w \mid \mathbf w \ge 0,\ \sum_i w_i = 1\}$. Equivalently, writing $\mathbf a_i$ for the $i$th column of $A$, the classical set can be expressed as the convex hull
$\mathcal C=\mathrm{conv}\{\mathbf a_1,\dots,\mathbf a_N\}$, since any admissible probability vector has the form $\mathbf p=\sum_i w_i\mathbf a_i$ with $w_i\ge0$ and $\sum_i w_i=1$.

To test whether an observed probability vector lies outside this classical set, we introduce a linear witness \cite{Morris2022,Greenwood2023,Gulati2024,Vandenberghe} specified by a coefficient vector $\mathbf c\in\mathbb R^{JK}$. The corresponding witness statistic is $S(\mathbf p)=\mathbf c\cdot\mathbf p$, having a maximal classical value 
\beq
\max_{\mathbf p\in\mathcal C}S(\mathbf p)=\max_{\mathbf w\ge0,\,\sum_i w_i=1}\sum_i w_i(\mathbf c\cdot\mathbf a_i)=\max_i \mathbf c\cdot\mathbf a_i\equiv S_{\rm cl}, 
\eeq
i.e. the maximum is attained at an extreme point of the convex polytope. We therefore quantify nonclassicality through the violation $\Delta(\mathbf c)=\mathbf c\cdot\mathbf p-S_{\rm cl}$, and define the optimal witness value $\Delta^\star=\max_{\|\mathbf c\|_2=1}\Delta(\mathbf c)$. A positive value $\Delta^\star>0$ certifies incompatibility with a single positive latent-variable description in the ideal noiseless setting.

The optimization over $\mathbf c$ can be cast explicitly as a convex program. Introducing an auxiliary scalar $t$ to represent the classical bound, one can rewrite
\[
\Delta^\star = \max_{\|\mathbf c\|_2 \le 1} \left( \mathbf c \cdot \mathbf p - \max_i \mathbf c \cdot \mathbf a_i \right)
\]
as the equivalent problem $\max_{\mathbf c,\, t} \quad \mathbf c \cdot \mathbf p - t$,  subject to $\mathbf c \cdot \mathbf a_i \le t \quad \forall i, \qquad \|\mathbf c\|_2 \le 1$.

This formulation makes clear that the search for an optimal witness amounts to finding a supporting hyperplane that maximally separates the observed statistics $\mathbf p$ from the classical polytope $\mathcal C$. The problem is convex and can be solved efficiently using standard numerical tools such as \texttt{CVXPY}. A detailed discussion of the numerical implementation, including computational scalability, runtime, and solver performance, is provided in Section 4.
We note that the nonclassicality test can also be formulated in a complementary manner as a feasibility or projection problem. Specifically, given observed statistics $\mathbf p$, one may seek a classical latent distribution $\mathbf w$ that best reproduces $\mathbf p$ by solving $\min_{\mathbf w} \;\|\!A\mathbf w - \mathbf p\|_2^2$ subject to $w_i \ge 0,\qquad \sum_i w_i = 1$. If $\mathbf p \in \mathcal C$, the minimum is zero and a feasible classical explanation exists. Otherwise, the optimal residual norm $\|A\mathbf w^\star - \mathbf p\|_2$ quantifies the distance of the observed statistics from the classical polytope. This provides an alternative measure of deviation from classical latent-variable consistency.

The witness-based formulation used in this work corresponds to the dual perspective, in which one seeks a hyperplane that maximally separates $\mathbf p$ from $\mathcal C$. The two viewpoints are closely related through convex duality: a positive witness violation implies a nonzero projection residual, while the latter provides a quantitative notion of distance to the classical set. In the present work we focus on the witness formulation, as it leads directly to experimentally accessible significance tests, but the projection-based approach offers a complementary perspective that may be useful in future studies.
\section{Statistical robustness under adversarial noise and finite sampling}
In the following, we fix a witness vector $\mathbf c$ attaining (or closely approximating) the optimum $\Delta^\star$, and we model systematic degradation of the mean statistics, analogous to reduced visibility in Bell tests \cite{Kwiat1993,Tomasin2017}. We also include additive physical noise. If ${\mathbf p}_q \in \mathbb R^{JK}$ denotes the ideal decoding statistics under test, assuming statistics incompatible with a single positive latent-variable model, we define the mean observed statistics as $\mathbf p_\alpha=(1-\alpha){\mathbf p}_q+\alpha{\mathbf p}_{\rm cl}$, where $\mathbf p_{\rm cl} \in \mathcal C$ is a classical distribution chosen to saturate the witness bound, and $\alpha \in [0,1]$. Thus $\alpha=0$ corresponds to the ideal statistics, while $\alpha=1$ yields a fully classical mean. 

We note that the choice of the classical admixture $\mathbf p_{\rm cl}$ entering the definition $\mathbf p_\alpha=(1-\alpha)\mathbf p_q+\alpha \mathbf p_{\rm cl}$ is taken to be adversarial, in the sense that $\mathbf p_{\rm cl}$ is chosen to saturate the witness bound and thus degrade the nonclassical signal as rapidly as possible. This construction provides a conservative estimate of robustness, analogous to worst-case visibility reductions in Bell-type tests. More generally, one may consider alternative admixture models of the form $\mathbf p_\alpha=(1-\alpha)\mathbf p_q+\alpha \tilde{\mathbf p}$, where $\tilde{\mathbf p}$ is an arbitrary classical or experimentally motivated noise distribution. In such cases, the witness value becomes $S(\mathbf p_\alpha)=(1-\alpha)\,S(\mathbf p_q)+\alpha\,S(\tilde{\mathbf p})$, and the rate at which the violation is reduced depends on the overlap of $\tilde{\mathbf p}$ with the optimal witness direction. Since $S(\tilde{\mathbf p}) \le S_{\mathrm{cl}}$ for any classical distribution, the adversarial choice used here yields the fastest possible degradation and therefore provides a lower bound on detectability. Other noise models, including isotropic or experimentally measured background distributions, will generally lead to slower degradation and hence improved robustness. Accordingly, the detection curves presented below should be interpreted as conservative benchmarks, rather than precise predictions for any specific experimental setting.

Furthermore, the probabilities $p(y_k\mid\theta_j)$ are estimated in practice from finite samples and therefore exhibit multinomial statistical fluctuations. The covariance structure is thus not isotropic, but instead takes the form
\[
\Sigma^{(j)}_{kk'}
=
\frac{1}{M_j}
\left[
p(y_k\mid\theta_j)\delta_{kk'}
-
p(y_k\mid\theta_j)p(y_{k'}\mid\theta_j)
\right],
\]
where $M_j$ is the number of experimental samples collected in measurement context $\theta_j$. The previous relation automatically enforces normalization within each measurement context. The corresponding witness statistic $S_{\rm obs}=\mathbf c\cdot\mathbf p_{\rm obs}$ has variance $\sigma_S^2=\mathbf c^{\mathsf T}\Sigma\mathbf c$. For sufficiently large sample sizes, the central limit theorem implies that the witness statistic is approximately Gaussian distributed. The corresponding detection probability is therefore approximated by
\[
P_{\mathrm{det}}(\alpha)=1-\Phi\!\left[
\frac{S_{\mathrm{cl}}+\kappa\sigma_S-\mu_\alpha}{\sigma_S}
\right].
\]
Here $\Phi(x)$ denotes the cumulative distribution function of the standard normal distribution, and $\mu_\alpha=\mathbf c\cdot\mathbf p_\alpha$ denotes the mean value of the witness statistic for the noisy distribution $\mathbf p_\alpha$, while $\kappa$ specifies the statistical confidence threshold in units of the standard deviation $\sigma_S$. For example, $\kappa=2$ corresponds to a one-sided Gaussian confidence level of approximately $97.7\%$.

In addition to this asymptotic Gaussian approximation, we explicitly validate the witness statistics using finite-sample Monte Carlo simulations based on multinomial sampling. Synthetic experimental counts are generated independently for each context and converted into empirical witness distributions, enabling direct comparison between exact multinomial statistics and the Gaussian approximation.
\begin{figure*}[th]
\begin{center}
\includegraphics[width=15 cm]{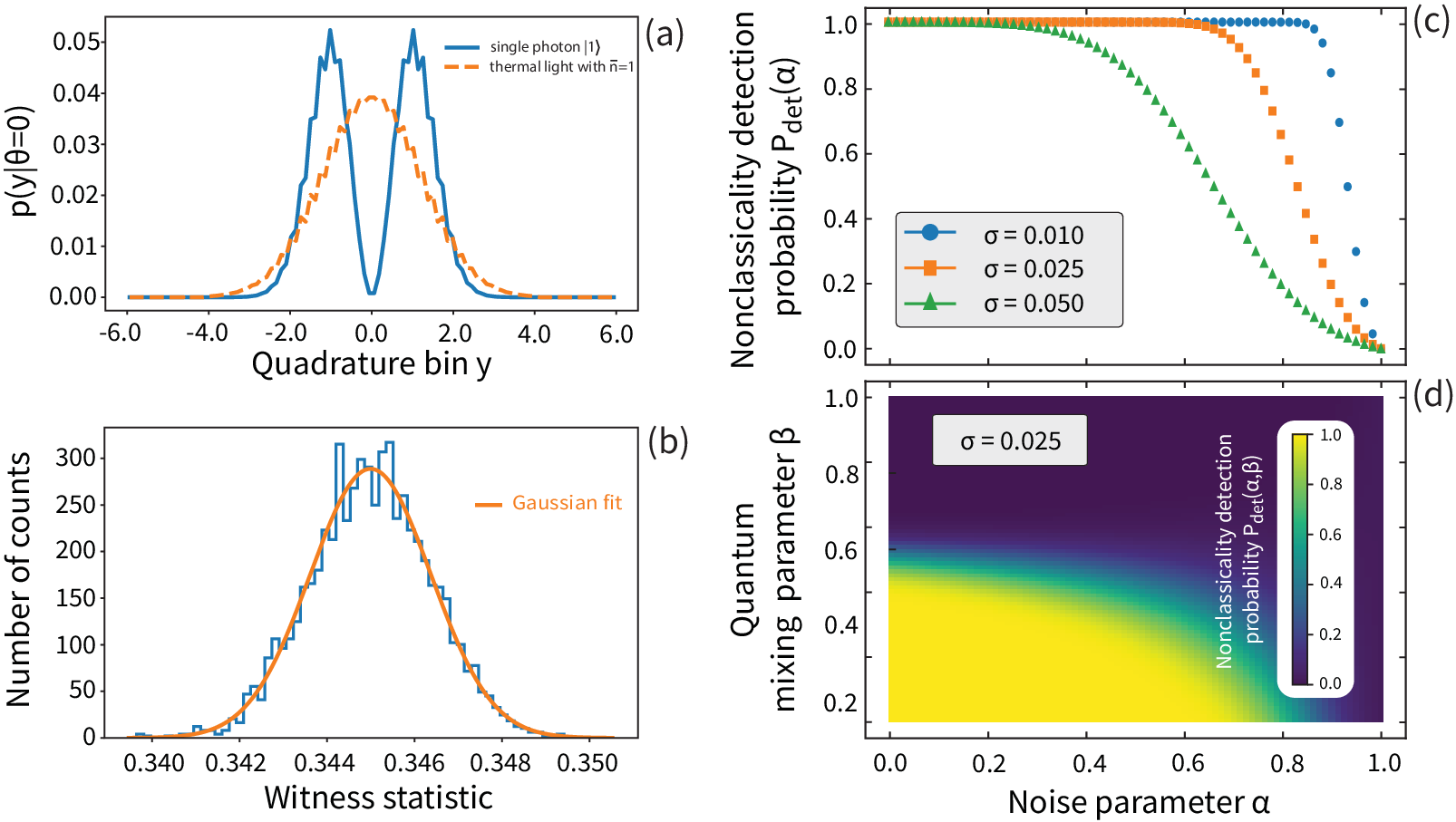}
\caption{
(a) Representative projected output distribution
$p(y|\theta=0)$
obtained from the discretized latent-space model for the single-photon state and for a thermal state with $\bar n=1$.
The horizontal axis denotes the continuous quadrature coordinate $y$, obtained from the projection
$y=\zeta\cos\theta+\eta\sin\theta$.
For the rotationally symmetric states considered here, the corresponding marginal distributions are theoretically independent of $\theta$, up to numerical discretization effects.
The double-peaked structure of the single-photon marginal reflects the nonclassical structure of the underlying Wigner function.
(b) Monte Carlo validation of multinomial sampling statistics for the witness observable.
Synthetic finite-sample measurements were generated from multinomial distributions associated with the projected probabilities $p(y_k|\theta_j)$.
The resulting histogram of witness values is well approximated by a Gaussian distribution, supporting the asymptotic central-limit approximation used in the analytical detection-probability model.
(c) Witness-detection probability as a function of the classical admixture parameter $\alpha$ for different effective Gaussian noise amplitudes $\sigma$.
The numerical implementation uses a two-dimensional latent phase space $(\zeta,\eta)$ discretized on a uniform $100\times100$ grid over the square $[-L,L]\times[-L,L]$ with $L=4$, yielding $N=10^4$ latent basis points.
The readout family consists of $J=25$ projection angles
$\theta_j=j\pi/J$.
For each context, the projected coordinate
$y=\zeta\cos\theta_j+\eta\sin\theta_j$
is discretized into $K=100$ uniformly spaced outcome bins over
$[-y_{\max},y_{\max}]$,
with
$y_{\max}=1.05\sqrt{2}\,L$.
This defines a forward matrix
$A\in\mathbb R^{2500\times10000}$.
(d) Detection-probability heat map
$P_{\rm det}(\alpha,\beta)$
as a function of the classical admixture parameter $\alpha$ and the quantum mixing parameter $\beta$ for fixed noise level $\sigma=0.025$.
The parameter $\beta$ interpolates between the single-photon state and a thermal state with $\bar n=1$.
The heat map quantifies the region in parameter space where violations of the single positive latent-variable model remain statistically detectable.
}\end{center}
\end{figure*}
\section{Numerical implementation and statistical validation}
For the numerical demonstration shown in Fig.~2, we use a two-dimensional latent variable $(\zeta,\eta)$ interrogated through a family of externally specified readout contexts $\theta$ that return linear projections
\[
y=\zeta\cos\theta+\eta\sin\theta.
\]
This construction is analogous to quantum homodyne tomography \cite{Raymer2009}, in which quadrature measurements at different angles correspond to Radon projections of an underlying phase-space quasiprobability distribution \cite{Schleich}. For each readout context $\theta_j$, the projected variable $y$ is discretized into outcome bins $\{y_k\}$, and the corresponding probabilities $p(y_k\mid\theta_j)$ are obtained by integrating the latent distribution over the associated regions in phase space.

The latent distribution is taken to be the Wigner function of the single-photon Fock state,
\[
W_{|1\rangle}(\zeta,\eta)
=
\frac{2}{\pi}
\left[
4(\zeta^2+\eta^2)-1
\right]
\exp\!\left[-2(\zeta^2+\eta^2)\right].
\]

To make the construction explicit, we briefly outline how the observable probabilities $p(y_k\mid\theta_j)$ are obtained from the Wigner function. For a given readout context $\theta$, the measured variable is the projection
\[
y=\zeta\cos\theta+\eta\sin\theta.
\]
The corresponding probability density is given by the marginal
\[
p(y\mid\theta)
=
\int d\zeta\, d\eta\;
W(\zeta,\eta)\,
\delta\!\left(y-\zeta\cos\theta-\eta\sin\theta\right),
\]
which corresponds to a Radon transform of the phase-space distribution. 

In practice, the outcomes are discretized into bins $\{y_k\}$, and the probabilities are obtained by integrating over each bin,
\[
p(y_k\mid\theta)
=
\int_{y\in \mathrm{bin}_k} dy\; p(y\mid\theta)
=
\int_{\mathcal R_{jk}} d\zeta\, d\eta\;
W(\zeta,\eta),
\]
where $\mathcal R_{jk}$ denotes the region of phase space corresponding to outcomes falling into bin $y_k$ for context $\theta_j$. An example of such a marginal is given in Fig. 2a.

In the numerical implementation, this integral is evaluated by discretizing the $(\zeta,\eta)$ plane on a uniform grid and summing the values of the Wigner function over the corresponding regions. This procedure yields the entries of the forward matrix $A$, with $A_{(j,k),i}$ representing the contribution of latent cell $i$ to outcome $y_k$ under context $\theta_j$. For a discretization with $N$ latent cells, $J$ measurement contexts, and $K$ output bins per context, the forward matrix has dimensions $A\in\mathbb R^{(JK)\times N}$. The computational cost of the witness optimization therefore scales primarily with the total number of observed statistics $JK$ and the latent-space resolution $N$. Increasing $N$ improves the approximation of the underlying continuous phase space, while increasing $J$ and $K$ refines the contextual and output resolution of the observable statistics. In the following Subsection we illustrate the corresponding trade-off between computational runtime and convergence of the witness margin as the latent discretization is refined.

The classical consistency problem and the associated witness optimization are solved using standard Python-based convex optimization tools implemented with \texttt{CVXPY}. To assess robustness at the level of quantum states and introduce a one-parameter family of latent quantum models with density operator
\[
\rho_\beta
=
(1-\beta)|1\rangle\langle1|
+
\beta\rho_{\mathrm{th}}(1),
\]
where
\[
\rho_{\mathrm{th}}(\bar n)
=
\sum_{n=0}^{\infty}
\frac{\bar n^n}{(\bar n+1)^{n+1}}
|n\rangle\langle n|
\]
denotes the thermal state with mean photon number $\bar n=1$. The mean photon number is therefore identical for all values of $\beta$, isolating changes in nonclassical structure from trivial variations in energy. A corresponding marginal distribution for the thermal state with $\bar n=1$ is also shown in Fig. 2a. The admixed Wigner function is $W_{\rho_\beta}=(1-\beta)W_{|1\rangle}+\beta W_{\rm th,1}$, where $W_{\rm th,1}=(2/3\pi)\exp\!\left[-2(\zeta^2+\eta^2)/3\right]$.

To calculate the detection probability, we use the Gaussian approximation discussed in the previous Section, whose validation is shown in Fig. 2b. The detection probability as a function of the noise parameter $\alpha$ is shown in Fig. 2c for three different values of the effective Gaussian noise parameter $\sigma$. Finally, the heat-map plot of Fig.~2(d) shows the detection probability as a function of both parameters $\alpha$ and $\beta$, demonstrating a substantial region within the quantum-classical transition and additional classical noise where non-classicality remains detectable.
\subsection{Finite-statistics robustness and computational scalability}
\begin{figure*}[th]
\begin{center}
\includegraphics[width=15 cm]{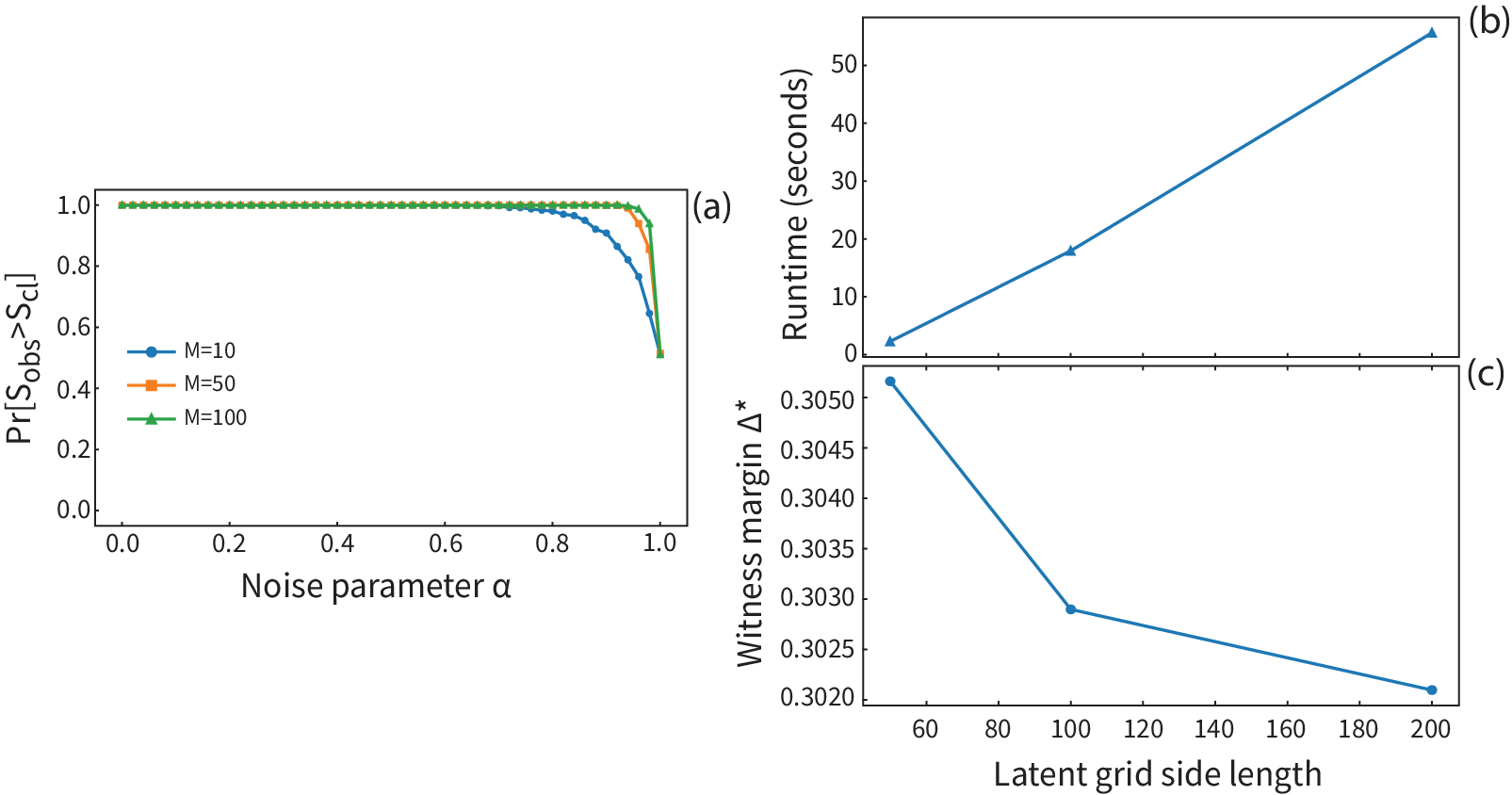}
\caption{
(a) Empirical probability that the sampled witness statistic exceeds the classical bound, obtained from explicit multinomial Monte Carlo simulations for different numbers of samples per measurement context,
$M=10,50,100$.
The curves display the probability that the sampled witness statistic exceeds the classical bound as a function of the classical admixture parameter $\alpha$.
Smaller values of $M$ produce broader statistical fluctuations and smoother degradation of detectability, while larger sample sizes approach the sharp asymptotic threshold behavior predicted analytically.
(b) Runtime of the witness optimization procedure as a function of latent-space grid size.
The observed scaling is consistent with the expected polynomial complexity of the convex optimization problem.
(c) Convergence of the optimized witness violation $\Delta^\star$ under refinement of the latent-space discretization.
The smooth convergence indicates that the observed witness violation is not a discretization artifact.
}
\end{center}
\end{figure*}
To test the statistical robustness of the witness framework, we performed explicit Monte Carlo simulations based on multinomial sampling. For each measurement context $\theta_j$, synthetic experimental counts were generated according to the multinomial distribution associated with the probabilities $p(y_k|\theta_j)$. The resulting empirical frequencies were assembled into the observed probability vector
$\mathbf p_{\rm obs}$, from which the witness statistic $S_{\rm obs}=\mathbf c\cdot\mathbf p_{\rm obs}$ was computed. The corresponding probability, $\Pr[S_{\rm obs}>S_{\rm cl}]$, is shown in Fig.~3a for different numbers of samples per context, $M=10,50,100$.
The results demonstrate that the witness remains highly robust against finite counting statistics. Smaller sample sizes produce broader multinomial fluctuations and therefore smoother degradation of detectability, whereas larger sample sizes rapidly approach the sharp asymptotic threshold behavior predicted by the Gaussian analysis. The comparison clarifies the distinction between the phenomenological Gaussian-noise model used in the analytical treatment and the exact finite-sampling statistics generated by multinomial counting noise alone.

Sparse representations of the forward matrix $A$ were used throughout the implementation. For the main calculation, the forward matrix has dimensions $A\in\mathbb R^{2500\times10000}$, but each latent cell contributes only to a small number of output bins across contexts. Exploiting this sparsity substantially reduces memory requirements and enables efficient convex optimization on a standard workstation.

Runtime measurements as a function of latent-space grid size are shown in Fig.~3b. The observed scaling is consistent with the expected polynomial complexity of the optimization procedure and demonstrates the computational feasibility of the method for the system sizes considered here.

Finally, we examined convergence with respect to the latent-space discretization. The main calculation uses a $100\times100$ discretization of the square $[-L,L]\times[-L,L]$, corresponding to $N=10^4$ latent cells. To test discretization dependence, the witness optimization was repeated for $50\times50$, $100\times100$, and $200\times200$ grids while keeping the number of measurement contexts and outcome bins fixed. As shown in Fig.~3c, the optimized witness violation $\Delta^\star$ converges smoothly under grid refinement, indicating that the separation from the classical polytope is not a discretization artifact.

Overall, these numerical tests show that the witness violation remains statistically robust under realistic finite-sampling conditions, converges stably under grid refinement, and can be computed efficiently for experimentally relevant latent-space discretizations.
\section{From phase-space variables to an effective spin system underlying latent space}
We now consider an effective spin system supposed to underlie the activation of a neuron. We stress that this discussion serves only as an illustration, and no claim is made regarding the physical realization of such degrees of freedom in neural systems. The purpose of this extension is twofold: first, to demonstrate that the proposed framework is not restricted to continuous phase spaces or special physical couplings between phase-space variables and observables, but applies equally to discrete quantum systems; and second, to establish a direct connection with binary or finite-outcome measurements, which are more representative of experimentally accessible readouts, including those encountered in neural recording contexts.

Specifically, we consider a spin-$j$ system prepared repeatedly in the same quantum state $\rho_j$. We will show that such a system can be mapped to the previous phase-space case. Now, each readout context $\theta$ specifies a direction $\boldsymbol n_\theta$ on the unit sphere $S^2$, and the corresponding observable is the spin projection $J_{\boldsymbol n_\theta}=\boldsymbol n_\theta\cdot\mathbf J$, with outcomes $\in\{-j,-j+1,\dots,j\}$. Rather than resolving individual spin outcomes, we use a binary, coarse-grained readout interpreted as a neuron activation: the outcome $y=H$ ("high activation") corresponds to spin projections exceeding a threshold, while $y=L$ (“low activation”) corresponds to projections below it. This defines a two-outcome POVM $\{E^{\rm H}_\theta,E^{\rm L}_\theta\}$, and the experimentally accessible statistic is the activation probability $p({\rm act}\mid\theta)=\mathrm{Tr}\!\left(\rho_jE^{\rm act}_\theta\right)$, with ${\rm act}={\rm H,L}$.

This construction admits a natural phase-space formulation without introducing an explicit decoder. The spin-$j$ system possesses an SU(2) phase space given by the sphere $S^2$, equipped with a Stratonovich-Weyl kernel $\Delta_j(\Omega)$, with $\Omega=(\vartheta,\varphi)$ \cite{Rundle2021}. Any state $\rho$ defines a spherical Wigner function $W_{\rho_j}(\Omega)=\mathrm{Tr}[\rho_j\Delta_j(\Omega)]$, while each activation POVM element admits a corresponding phase-space symbol $\Xi^{\rm act}_\theta(\Omega)=\mathrm{Tr}[E^{\rm act}_\theta\,\Delta_j(\Omega)]$. The observed activation probabilities can therefore be written as $p({\rm act}\mid\theta)=\int d\Omega\, W_{\rho_j}(\Omega)\,\Xi^{\rm act}_\theta(\Omega)$, which is formally identical to the latent phase-space representation introduced earlier. We note that while the formal phase-space construction applies to all spins, it is only for $j> \tfrac{1}{2}$ that the outcome space and contextual incompatibility lead to meaningful restrictions on latent representations. This mirrors the fact that contextuality in the sense of Kochen–Specker arises only in Hilbert spaces of dimension three or higher \cite{Gleason,KS}. Interestingly, in the large-$j$ limit, the spin-$j$ formulation reduces locally to an effective oscillator description via the Holstein--Primakoff approximation \cite{Polzik2010}, recovering the planar phase space and quadrature variables familiar from the earlier Wigner-based example. This observation is particularly relevant in quantum biophysics, where vibronic degrees of freedom are center stage, as for example in photosynthetic light-harvesting \cite{Ishizaki2009,Scholes2017,Cao2020}. 

More generally, this construction shows that the latent-variable consistency constraints developed in the present framework are not restricted to continuous phase-space descriptions, but naturally extend to finite-dimensional contextuality settings familiar from quantum foundations. To better clarify the spin-$j$ construction, we consider a concrete realization for $j=1$ (a qutrit), based on the KCBS pentagon construction. We define five rank-one projectors $P_i = |v_i\rangle\langle v_i|$, $i=1,\dots,5$, arranged such that adjacent projectors are orthogonal, $P_i P_{i+1}=0$ (with indices modulo five). A convenient parametrization is
\[
|v_i\rangle =
\left(
\cos\chi,\,
\sin\chi\cos\phi_i,\,
\sin\chi\sin\phi_i
\right),
\qquad
\phi_i = \frac{4\pi i}{5},
\]
with $\cos^2\chi = 1/\sqrt{5}$. One readily verifies that $\langle v_i | v_{i+1} \rangle=\cos^2\chi + \sin^2\chi \cos(\phi_i-\phi_{i+1}) = 0$, so that neighboring projectors are mutually exclusive. For the state $|\psi\rangle=(1,0,0)$, the corresponding probabilities are
\[
p_i = \langle \psi|P_i|\psi\rangle = |\langle \psi|v_i\rangle|^2 = \cos^2\chi = \frac{1}{\sqrt{5}},
\]
and hence $\sum_{i=1}^5 p_i = \sqrt{5}$. In any noncontextual latent-variable model with a single positive distribution, one assigns predetermined values $v_i\in\{0,1\}$ subject to exclusivity constraints $v_i+v_{i+1}\le 1$, implying $\sum_i v_i\le 2$ and therefore the classical bound $\sum_i p_i \le 2$. The quantum value $\sqrt{5}>2$ thus demonstrates that no single positive latent distribution can reproduce the observed statistics across all contexts.

The violation derived above should thus be interpreted as the impossibility of rewriting the observed statistics in the form
\[
p({\rm act}\mid\theta)
=
\int d\Omega\, p(\Omega)\, p({\rm act}\mid\Omega,\theta),
\qquad
p(\Omega)\ge 0,\quad p({\rm act}\mid\Omega,\theta)\ge 0,
\]
with a single underlying positive distribution. In this sense, nonclassicality does not hinge on negativity of a particular representation (such as the Wigner function) alone. Rather, it reflects the more general statement that no representation exists in which both the state and the measurement response functions can be simultaneously interpreted as positive probabilities across all contexts. The phase-space representation should therefore be regarded only as a convenient linear parametrization of the statistics. The essential result is that no underlying model with simultaneously positive latent distributions and positive response functions can reproduce the observed contextual statistics.
\section{Towards experimental realization}
Although autoencoders capture general principles of efficient compression and reconstruction that are widely believed to be relevant for neural information processing, the formalism developed here does not assume that neural systems implement autoencoders or explicit latent representations. Instead, the framework is formulated entirely in terms of observable decoding statistics obtained under multiple contexts. Once abstracted from this motivating example, the proposed tests apply directly to experimental scenarios accessible with modern neural recording and control technologies. 

High-density electrode arrays enable simultaneous recording of large neural populations with millisecond temporal resolution, allowing decoding statistics to be estimated with high precision across repeated trials, with statistical uncertainties scaling as $1/\sqrt{M}$ in the number of trials $M$ \cite{Averbeck2006,Paninski2010,Stevenson2011}. Optogenetic stimulation provides a flexible means to define and manipulate readout contexts by selectively perturbing subpopulations, modulating network states, or implementing closed-loop feedback protocols \cite{Grosenick2015}. In vitro neuronal cultures and brain organoids \cite{Trujillo2019Organoids} grown on multielectrode arrays offer high-throughput platforms with precise stimulation control and long recording stability, enabling the systematic collection of large datasets across many contexts \cite{Sharf2019,Jun2017}. Assuming only statistical errors, we can connect the number of trials $M$ with the standard deviation $\sigma$ used in our numerical demonstration, noting that for roughly uniform binning with $p(y_k\mid\theta_j)\approx 1/K$ one obtains $\sigma\sim 1/\sqrt{KM}$. With the binning used in our numerical implementation ($K=100$) and $M=100$ trials per context, this gives $\sigma\sim 1/\sqrt{10^4}=0.01$, matching the noise levels explored in our detection curves. Such trial numbers are achievable with the aforementioned techniques. A schematic of a possible experimental realization is shown in Fig. 4.
\begin{figure*}[th]
\begin{center}
\includegraphics[width=12 cm]{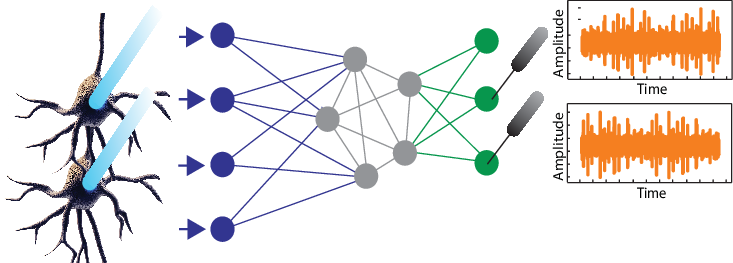}
\caption{Possible experimental realization of the proposed paradigm. Optogenetic stimulation and electrode recordings provide access to population-level activity and enable variation of decoding contexts.}
\end{center}
\end{figure*}
\section{Discussion}
The framework introduced in this work should primarily be interpreted as a model-agnostic statistical consistency test for autoencoder latent-variable descriptions across multiple readout contexts. A violation of the proposed witness indicates that the observed statistics cannot be reproduced by a single positive latent-variable distribution satisfying the assumptions of the framework. Such a violation, however, does not by itself establish the presence of microscopic quantum coherence or entanglement in neural systems, nor does it identify a specific underlying physical mechanism. 

Unlike standard contextuality frameworks, the construction developed here does not assume a predefined Hilbert-space structure, compatibility graph, or measurement algebra. Instead, it is formulated directly in terms of conditional distributions inferred from data and decoder-dependent readout statistics. The novelty of the approach therefore lies not in introducing a fundamentally new notion of contextuality, but in reformulating contextuality-inspired consistency tests in a model-agnostic latent-representation setting motivated by machine learning and neural information processing.

We also note that a variety of well-established notions of nonclassicality and contextuality exist in the quantum foundations literature, including Bell-type inequalities such as CHSH, Kochen--Specker contextuality tests such as KCBS, contextual fraction, and quasiprobability-based measures such as Wigner negativity and its associated robustness measures. These approaches typically rely on a specified measurement structure and an underlying operational or physical model. The framework developed here is complementary in scope and intent. Rather than benchmarking or optimizing known contextuality inequalities, our goal is to provide a data-driven test of classical latent-variable consistency directly at the level of observable statistics. Establishing quantitative relations between the present framework and existing measures of contextuality or quasiprobability negativity would require embedding the observed statistics into a fully specified operational theory, which lies beyond the scope of the present work.

At present, the proposed witness should be viewed as a structural exclusion test rather than a complete operational resource theory. A violation establishes that the observed statistics cannot be reproduced by a single positive latent-variable model within the assumptions of the framework, but it does not by itself imply a specific operational advantage. In particular, the present work does not establish quantitative relations between witness violations and properties such as compression efficiency, reconstruction fidelity, or energetic cost. Exploring whether such relations exist in realistic learning systems constitutes an interesting direction for future work. 

The proposed Bell-type construction relies on two important assumptions: (i) that changes in measurement context do not modify the underlying latent distribution, and (ii) that the latent state remains sufficiently stable over the timescale of data acquisition. In realistic neural systems, these assumptions may fail, for example if stimulation protocols perturb the underlying state or induce slow temporal drift. Such effects may also arise from model misspecification, insufficient latent dimensionality, or experimental nonstationarity, and can in principle generate apparent witness violations even within a classical framework. At the level of the witness, nonstationary perturbations may be modeled as fluctuations $\mathbf p \to \mathbf p + \delta \mathbf p(t)$, producing corresponding fluctuations in the observed witness value $S_{\rm obs} = \mathbf c \cdot \mathbf p_{\rm obs}$ of order $\mathbf c \cdot \delta \mathbf p$. Distinguishing genuine contextuality-like effects from such classical mechanisms therefore requires careful experimental design, including randomized or interleaved presentation of measurement contexts, explicit stationarity tests, and comparison of witness values across temporally separated subsets of trials. 

Several important directions for future work remain open. One natural extension is the application of the present framework to biologically realistic or experimentally obtained datasets, including spiking neural recordings, recurrent neural-network dynamics, or latent representations learned by artificial neural networks. Implementing such tests requires defining suitable measurement contexts, estimating conditional output distributions across repeated trials, and controlling for nonstationarity and noise. Another important direction is the application of the framework to trained autoencoder models. In particular, it would be instructive to analyze classical architectures such as variational autoencoders, for which one expects $\Delta^\star \approx 0$, as well as quantum autoencoders, which may exhibit $\Delta^\star > 0$. Such studies would provide a concrete benchmark for assessing the ability of the proposed witness to distinguish classical from genuinely quantum compression mechanisms in practical information-processing settings.

In summary, we have introduced a model-agnostic framework for testing nonclassical statistics in neural representations, using autoencoders as a transparent motivating example. Our work shifts the search for quantum-like signatures in neural function away from microscopic mechanisms and toward robust, ensemble-level constraints on information processing. The perspective put forward in this work suggests that tools developed to test the foundations of quantum physics may also illuminate the fundamental physics of biological intelligence.

%
%

\section{Acknowledgements}
C. X. acknowledges the National Natural Science Foundation of China (grant number T235005). S. L. acknowledges the National Natural Science Foundation of China (grant number U21A20437). G.P.T. acknowledges the Department of Navy award N629092412119 issued by the Office of Naval Research Global, USA. M. S. acknowledges project PID2024-162141OB-I00, funded by MICIU/AEI/10.13039/501100011033/ FEDER, UE, Ayuda Ram\'on y Cajal 2021 (RYC2021-032032-I,
MICIU/AEI/10.13039/501100011033, ESF+) as well as Project FEDER C-EXP-256-UGR23 Consejer\'ia de Universidad, Investigaci\'on e Innovaci\'on y UE Programa FEDER Andaluc\'ia 2021-2027.


\begin{thebibliography}{99}

\bibitem{Hameroff1996}
Hameroff S and Penrose R 1996 Orchestrated reduction of quantum coherence in brain microtubules: A model for consciousness {\it Math. Comput. Simul.} {\bf 40} 453

\bibitem{Tegmark2000}
Tegmark M 2000 Importance of quantum decoherence in brain processes {\it Phys. Rev. E} {\bf 61} 4194

\bibitem{Hagan2002}
Hagan S, Hameroff S R and Tuszyński J A 2002 Quantum computation in brain microtubules: Decoherence and biological feasibility {\it Phys. Rev. E} {\bf 65} 061901

\bibitem{Adams2020}
Adams B and Petruccione F 2020 Quantum effects in the brain: A review {\it AVS Quantum Sci.} {\bf 2} 022901

\bibitem{Gassab2025}
Gassab L, Pusuluk O, Cattaneo M and Müstecaplıoğlu Ö E 2025 Quantum models of consciousness from a quantum information science perspective {\it Entropy} {\bf 27} 243

\bibitem{Scholes2017}
Scholes G D {\it et al.} 2017 Using coherence to enhance function in chemical and biophysical systems {\it Nature} {\bf 543} 647

\bibitem{Kominis2015}
Kominis I K 2015 The radical-pair mechanism as a paradigm for the emerging science of quantum biology {\it Mod. Phys. Lett. B} {\bf 29} 1530013

\bibitem{Kominis2025}
Kominis I K and Gkoudinakis E 2025 Approaching the quantum limit of energy resolution in animal magnetoreception {\it PRX Life} {\bf 3} 013004

\bibitem{Plenio2013}
Huelga S F and Plenio M B 2013 Vibrations, quanta and biology {\it Contemp. Phys.} {\bf 54} 181

\bibitem{Kominis2020}
Kominis I K 2020 Quantum relative entropy shows singlet--triplet coherence is a resource in the radical-pair mechanism of biological magnetic sensing {\it Phys. Rev. Res.} {\bf 2} 023206

\bibitem{Tegmark2015}
Tegmark M 2015 Consciousness as a state of matter {\it Chaos Solitons Fractals} {\bf 76} 238

\bibitem{Tononi2016}
Tononi G, Boly M, Massimini M and Koch C 2016 Integrated information theory: from consciousness to its physical substrate {\it Nat. Rev. Neurosci.} {\bf 17} 450

\bibitem{Georgiev2025}
Georgiev D D 2025 Quantum information theoretic approach to the hard problem of consciousness {\it Biosystems} {\bf 251} 105458

\bibitem{Tuszynski2026}
Theise N D and Tuszynski J A 2026 Non-linearity, complexity, and quantization concepts in biology {\it Front. Hum. Neurosci.} {\bf 19} 1695510

\bibitem{Fisher2015}
Fisher M P A 2015 Quantum cognition: The possibility of processing with nuclear spins in the brain {\it Ann. Phys.} {\bf 362} 593

\bibitem{Halpern2019}
Yunger Halpern N and Crosson E 2019 Quantum information in the Posner model of quantum cognition {\it Ann. Phys.} {\bf 407} 92

\bibitem{Rourk2021}
Rourk C, Huang Y, Chen M and Shen C 2021 Indication of strongly correlated electron transport and Mott insulator in disordered multilayer ferritin structures {\it Materials} {\bf 14} 4527

\bibitem{Manzano2022}
Torres J J and Manzano D 2022 A model of interacting quantum neurons with a dynamic synapse {\it New J. Phys.} {\bf 24} 073007

\bibitem{Liu2024}
Liu Z, Chen Y-C and Ao P 2024 Entangled biphoton generation in the myelin sheath {\it Phys. Rev. E} {\bf 110} 024402

\bibitem{Neven2024}
Neven H {\it et al.} 2024 Testing the conjecture that quantum processes create conscious experience {\it Entropy} {\bf 26} 460

\bibitem{Manzano2024}
Torres J J and Manzano D 2024 Dissipative quantum Hopfield network: a numerical analysis {\it New J. Phys.} {\bf 26} 103018

\bibitem{Carleo2019}
Carleo G {\it et al.} 2019 Machine learning and the physical sciences {\it Rev. Mod. Phys.} {\bf 91} 045002

\bibitem{Mehta2019}
Mehta P {\it et al.} 2019 A high-bias, low-variance introduction to machine learning for physicists {\it Phys. Rep.} {\bf 810} 1

\bibitem{TsironisBook}
Tsironis G 2025 {\it Artificial intelligence and complex dynamical systems} (Springer Nature)

\bibitem{Petruccione2015}
Schuld M, Sinayskiy I and Petruccione F 2015 An introduction to quantum machine learning {\it Contemp. Phys.} {\bf 56} 172

\bibitem{Biamonte2017}
Biamonte J {\it et al.} 2017 Quantum machine learning {\it Nature} {\bf 549} 195

\bibitem{Briegel2018}
Dunjko V and Briegel H J 2018 Machine learning and artificial intelligence in the quantum domain {\it Rep. Prog. Phys.} {\bf 81} 074001

\bibitem{Lennie2003}
Lennie P 2003 The cost of cortical computation {\it Curr. Biol.} {\bf 13} 493

\bibitem{Baldi2012}
Baldi P 2012 Autoencoders, unsupervised learning, and deep architectures {\it JMLR Workshop Proc.} {\bf 27} 37

\bibitem{Bengio2013}
Bengio Y, Courville A and Vincent P 2013 Representation learning {\it IEEE Trans. Pattern Anal. Mach. Intell.} {\bf 35} 1798

\bibitem{Michelucci2022}
Michelucci U 2022 An introduction to autoencoders {\it arXiv:2201.03898}

\bibitem{Giryes2023}
Bank D, Koenigstein N and Giryes R 2023 Autoencoders in Machine Learning for Data Science Handbook pp 353--374

\bibitem{Tishby2015}
Tishby N and Zaslavsky N 2015 Deep learning and the information bottleneck principle {\it IEEE ITW} pp 1--5

\bibitem{Parrondo2015}
Parrondo J M R, Horowitz J M and Sagawa T 2015 Thermodynamics of information {\it Nat. Phys.} {\bf 11} 131

\bibitem{Anders2016}
Vinjanampathy S and Anders J 2016 Quantum thermodynamics {\it Contemp. Phys.} {\bf 57} 545

\bibitem{Spekkens2008}
Spekkens R W 2008 Negativity and contextuality are equivalent notions of nonclassicality {\it Phys. Rev. Lett.} {\bf 101} 020401

\bibitem{Cabello2008}
Cabello A 2008 Experimentally testable state-independent quantum contextuality {\it Phys. Rev. Lett.} {\bf 101} 210401

\bibitem{Horodecki2009}
Horodecki R {\it et al.} 2009 Quantum entanglement {\it Rev. Mod. Phys.} {\bf 81} 865

\bibitem{Brunner2014}
Brunner N {\it et al.} 2014 Bell nonlocality {\it Rev. Mod. Phys.} {\bf 86} 419

\bibitem{Shalm2015}
Shalm L K {\it et al.} 2015 Strong loophole-free test of local realism {\it Phys. Rev. Lett.} {\bf 115} 250402

\bibitem{Cavalcanti2017}
Cavalcanti D and Skrzypczyk P 2017 Quantum steering {\it Rep. Prog. Phys.} {\bf 80} 024001

\bibitem{Watts2021}
Watts A B, Yunger Halpern N and Harrow A 2021 Nonlinear Bell inequality {\it Phys. Rev. A} {\bf 103} L010202

\bibitem{Budroni2022}
Budroni C {\it et al.} 2022 Kochen--Specker contextuality {\it Rev. Mod. Phys.} {\bf 94} 045007

\bibitem{Wilde2010}
Wilde M M, McCracken J M and Mizel 2010 A Could light harvesting complexes exhibit non-classical effects at room temperature? {\it Proc. A } {\bf 466} 1347

\bibitem{YungerHalpern2020}
Yunger Halpern and Limmer D T 2020 Fundamental limitations on photoisomerization from thermodynamic resource theories  {\it Phys. Rev. A} {\bf 101} 042116

\bibitem{Spaventa2025}
Tiwary S, Spaventa G, Huelga S F and Plenio M B 2025 Quantum-resource-theoretical analysis of the role of vibrational structure in photoisomerization  {\it Phys. Rev. A} {\bf 112} 032440

\bibitem{Guzik2017}
Romero J, Olson J P and Aspuru-Guzik A 2017 Quantum autoencoders for efficient compression of quantum data {\it Quantum Sci. Technol.} {\bf 2} 045001

\bibitem{Prieto2021}
Bravo-Prieto C 2021 Quantum autoencoders with enhanced data encoding {\it Mach. Learn.: Sci. Technol.} {\bf 2} 035028

\bibitem{Sakhnenko2022}
Sakhnenko A, O'Meara C, Ghosh K J B, Mendl C B, Cortiana G and Bernab\'{e}-Moreno J 2022 Hybrid classical--quantum autoencoder for anomaly detection {\it Quantum Mach. Intell.} {\bf 4} 27

\bibitem{Locher2023}
Locher D F, Cardarelli L and M\"{u}ller M 2023 Quantum error correction with quantum autoencoders {\it Quantum} {\bf 7} 942

\bibitem{Tsironis2023}
Chalkiadakis A, Theocharakis M, Barmparis G D and Tsironis G P 2023 Quantum neural networks for the discovery and implementation of quantum error-correcting codes {\it Chaos} {\bf 33} 113127

\bibitem{Ma2024}
Ma H, Huang C-J, Chen C, Dong D, Wang Y, Wu R-B and Xiang G-Y 2023 On compression rate of quantum autoencoders: Control design, numerical and experimental realization {\it Automatica} {\bf 147} 110659

\bibitem{Wu2025}
Wu J, Fu H, Zhu M, Zhang H, Xie W and Li X-Y 2024 Quantum circuit autoencoder {\it Phys. Rev. A} {\bf 109} 032623

\bibitem{Frehner2025}
Frehner R and Stockinger K 2025 Applying quantum autoencoders for time series anomaly detection {\it Quantum Mach. Intell.} {\bf 7} 59

\bibitem{AAE}
Makhzani A, Shlens J, Jaitly N, Goodfellow I and Frey B 2015 Adversarial autoencoders {\it arXiv:1511.05644}

\bibitem{Morris2022}
Morris B, Fiderer L J, Lang B and Goldwater D 2022 Witnessing Bell violations through probabilistic negativity {\it Phys. Rev. A} {\bf 105} 032202

\bibitem{Greenwood2023}
Greenwood A C B, Wu L T H, Zhu E Y, Kirby B T and Qian L 2023 Machine-learning-derived entanglement witnesses {\it Phys. Rev. Appl.} {\bf 19} 034058

\bibitem{Gulati2024}
Gulati V, Singh G and Dorai K 2024 Using linear and nonlinear entanglement witnesses to generate and detect bound entangled states on an IBM quantum processor {\it Phys. Scr.} {\bf 99} 115122

\bibitem{Vandenberghe}
Boyd S and Vandenberghe L 2004 {\it Convex Optimization} (Cambridge: Cambridge University Press)

\bibitem{Efron1979}
Efron B 1979 Bootstrap methods: Another look at the jackknife {\it Ann. Stat.} {\bf 7} 1

\bibitem{EfronTibshirani1993}
Efron B and Tibshirani R J 1994 {\it An Introduction to the Bootstrap} (New York: Chapman \& Hall/CRC)

\bibitem{Larsson1998}
Larsson J-\AA{} 1998 Bell's inequality and detector inefficiency {\it Phys. Rev. A} {\bf 57} 3304

\bibitem{Gill2014}
Gill R D 2014 Statistics, causality and Bell's theorem {\it Stat. Sci.} {\bf 29} 512

\bibitem{Kwiat1993}
Kwiat P G, Steinberg A M and Chiao R Y 1993 High-visibility interference in a Bell-inequality experiment for energy and time {\it Phys. Rev. A} {\bf 47} R2472

\bibitem{Tomasin2017}
Tomasin M, Mantoan E, Jogenfors J, Vallone G, Larsson J-\AA{} and Villoresi P 2017 High-visibility time-bin entanglement for testing chained Bell inequalities {\it Phys. Rev. A} {\bf 95} 032107

\bibitem{Raymer2009}
Lvovsky A I and Raymer M G 2009 Continuous-variable optical quantum-state tomography {\it Rev. Mod. Phys.} {\bf 81} 299

\bibitem{Schleich}
Schleich W P 2001 {\it Quantum Optics in Phase Space} (Berlin: Wiley-VCH)

\bibitem{Rundle2021}
Rundle R P and Everitt M J 2021 Overview of the phase space formulation of quantum mechanics with application to quantum technologies {\it Adv. Quantum Technol.} {\bf 4} 2100016

\bibitem{Gleason}
Gleason A M 1957 Measures on the closed subspaces of a Hilbert space {\it J. Math. Mech.} {\bf 6} 885

\bibitem{KS}
Kochen S and Specker E P 1967 The problem of hidden variables in quantum mechanics {\it J. Math. Mech.} {\bf 17} 59

\bibitem{Polzik2010}
Hammerer K, S{\o}rensen A S and Polzik E S 2010 Quantum interface between light and atomic ensembles {\it Rev. Mod. Phys.} {\bf 82} 1041

\bibitem{Ishizaki2009}
Ishizaki A and Fleming G R 2009 Theoretical examination of quantum coherence in a photosynthetic system at physiological temperature {\it Proc. Natl Acad. Sci. USA} {\bf 106} 17255

\bibitem{Cao2020}
Cao J {\it et al.} 2020 Quantum biology revisited {\it Sci. Adv.} {\bf 6} eaaz4888

\bibitem{Averbeck2006}
Averbeck B B, Latham P E and Pouget A 2006 Neural correlations, population coding and computation {\it Nat. Rev. Neurosci.} {\bf 7} 358

\bibitem{Paninski2010}
Paninski L, Pillow J and Lewi J 2007 Statistical models for neural encoding, decoding, and optimal stimulus design {\it Prog. Brain Res.} {\bf 165} 493

\bibitem{Stevenson2011}
Stevenson I H and Kording K P 2011 How advances in neural recording affect data analysis {\it Nat. Neurosci.} {\bf 14} 139

\bibitem{Grosenick2015}
Grosenick L, Marshel J H and Deisseroth K 2015 Closed-loop and activity-guided optogenetic control {\it Neuron} {\bf 86} 106

\bibitem{Jun2017}
Jun J J {\it et al.} 2017 Fully integrated silicon probes for high-density recording of neural activity {\it Nature} {\bf 551} 232

\bibitem{Sharf2019}
Sharf T {\it et al.} 2022 Functional neuronal circuitry and oscillatory dynamics in human brain organoids {\it Nat. Commun.} {\bf 13} 4403

\bibitem{Obien2015MEA}
Obien M E J, Deligkaris K, Bullmann T, Bakkum D J and Frey U 2015 Revealing neuronal function through microelectrode array recordings {\it Front. Neurosci.} {\bf 8} 423

\bibitem{Trujillo2019Organoids}
Trujillo C A {\it et al.} 2019 Complex oscillatory waves emerging from cortical organoids model early human brain network development {\it Cell Stem Cell} {\bf 25} 558

\bibitem{Packer2015Optogenetics}
Packer A M, Russell L E, Dalgleish H W P and H{\"a}usser M 2015 Simultaneous all-optical manipulation and recording of neural circuit activity with cellular resolution in vivo {\it Nat. Methods} {\bf 12} 140

\bibitem{Stringer2019Manifolds}
Stringer C, Pachitariu M, Steinmetz N, Reddy C, Carandini M and Harris K D 2019 High-dimensional geometry of population responses in visual cortex {\it Nature} {\bf 571} 361
\end{thebibliography}
\end{document}